\documentclass[10pt]{article}
\usepackage{amsmath}
\usepackage{cite} 
\usepackage{indentfirst}
\usepackage{latexsym}
\usepackage{pstricks}

\long\def\ca#1\cb{} 

\newcommand{\ad}{^\dagger }

\newcommand{\AND}{{\small AND}}

\newcommand{\becs}{\begin{cases}}
\newcommand{\bem}{\begin{matrix}}
\newcommand{\blp}{\bigl(}

\newcommand{\brp}{\bigr)}
\newcommand{\bsk}{\bigskip }

\newcommand{\dya}[1]{|#1\rangle\langle#1|}

\newcommand{\encs}{\end{cases}}
\newcommand{\enm}{\end{matrix}}

\newcommand{\ket}[1]{|#1\rangle }

\newcommand{\od}{\odot }

\newcommand{\ot}{\otimes }

\newcommand{\st}{\sqrt{2}}

\newcommand{\Tr}{{\rm Tr}}

\newcommand{\vb}{\,|\,}

\newcommand{\AC}{{\mathcal A}}

\newcommand{\HC}{{\mathcal H}}

\newcommand{\al}{\alpha }
\newcommand{\bt}{\beta }

\newcommand{\dl}{\delta }

\newcommand{\lm}{\lambda }

\newcommand{\sg}{\sigma }

\begin{document}

\title{Hilbert Space Quantum Mechanics Is Noncontextual}

\author{Robert B. Griffiths
\thanks{Electronic mail: rgrif@cmu.edu}\\ 
Department of Physics,
Carnegie-Mellon University,\\
Pittsburgh, PA 15213, USA}
\date{Version of 16 October 2012}
\maketitle  


\begin{abstract}
  It is shown that quantum mechanics is noncontextual if quantum properties
  are represented by subspaces of the quantum Hilbert space (as proposed by
  von Neumann) rather than by hidden variables.  In particular, a measurement
  using an appropriately constructed apparatus can be shown to reveal the
  value of an observable $A$ possessed by the measured system before the
  measurement took place, whatever other compatible ($[B,A]=0$) observable $B$
  may be measured at the same time.
\end{abstract}

\bsk

	\section{Introduction}
\label{sct1}

The question of whether quantum mechanics is in some sense ``contextual''
seems to have first been raised by Bell, Sec.~5 of \cite{Bll66}, in the course
of a discussion of the possible form which might be taken by hidden variables
in quantum theory, i.e., variables which are in addition to the Hilbert space
customarily used to discussing quantum mechanics.  The basic idea, not
altogether easy to extract from Bell's paper, is the following (Ch.~7 of
\cite{Prs93} or Sec.~VII of \cite{Mrmn93}): Let $A$, $B$, $C$ be three
observables, represented by Hermitian operators on a finite-dimensional
Hilbert space, for some system which is to be measured, and suppose that $A$
commutes with $B$ and with $C$, but $B$ and $C$ do not commute with each
other.  Then according to standard quantum mechanics it is possible in
principle to measure $A$ and $B$ together in a single experiment, and it is
possible to measure $A$ and $C$ together, though all three cannot be measured
simultaneously. Would a measurement of $A$ when carried out along with a
measurement of $B$ yield the same result (for $A$) as a measurement of $A$
carried out along with a measurement of $C$?%
\footnote{The term ``contextual'' as used here should not be confused with the
  idea of a ``dependent'' or ``contextual'' event as employed in Ch.~14 of
\cite{Grff02c}.  The latter makes no reference to measurements, and is a general
feature of Hilbert space quantum mechanics when applied to multiple systems
described by a tensor product.}

The answer would seem to be an obvious ``Yes'' if the measurement, however it
is carried out, reveals a value of $A$ that was there before the measurement
took place.  We shall argue that this is, indeed, correct if the apparatus has
been designed and operated by a competent experimentalist.  However, a
substantial literature has accumulated which would throw doubt on this or
suggest the opposite on the basis of various arguments related to the
Kochen-Specker%
\footnote{As essentially the same result was obtained and published slightly
  earlier by Bell in \cite{Bll66}, it would be appropriate to refer to
  the ``Bell-Kochen-Specker'' theorem. However, in the rest of this paper we
  follow the common usage in the philosophical literature.}  
theorem \cite{KcSp67}.  A recent discussion by Hermens \cite{Hrmn11} provides
an extensive list of references to work published since 2000, to which could be
added much additional material that has appeared in the last two years.  How
have so many come to this conclusion?  By adopting, we shall argue, the view
that the real world is \emph{classical}, contrary to all we have learned from
the development of quantum mechanics in the twentieth century.  In particular,
discussions couched in terms of hidden variables typically assume that they are
classical rather than the sort of thing one might expect in a quantum
mechanical world.

Such confusion is not unrelated to the failure of introductory quantum
mechanics textbooks and courses to provide a proper discussion of what actually
happens during a quantum measuring process.  Rather than explaining how a piece
of apparatus interacts with and thereby extracts and amplifies information
about the measured system, students are given calculational tools, such as wave
function collapse, whose connection with the quantities supposedly measured is
at best vague and sometimes quite misleading.  When Bell introduced the idea of
contextuality in 1966 there was no satisfactory theory of quantum measurement
processes, and so it was reasonable to ask whether measurement outcomes
revealed values of some sort of hidden variable(s). But developments since
then, beginning in the 1980s \cite{Grff84,Omns88,GMHr90}, allow measurements
and other quantum processes to be described in a fully consistent way using
only the quantum Hilbert space along with a stochastic quantum dynamics based
upon extensions of Born's rule. A very detailed presentation of this
(consistent or decoherent) histories approach will be found in \cite{Grff02c},
and shorter treatments in \cite{Hhnb10,Grff11b,Grff09b}. Various conceptual
issues and criticisms are discussed in \cite{Grff12}.

The histories approach is not the only interpretation of quantum mechanics to
employ the Hilbert space without any additional hidden variables to represent
physical properties. The alternatives include the Everett or ``many worlds''
interpretation \cite{Evrt57,Brrt09,Vdmn09,Brrt10}, various modal
interpretations \cite{Hmmo09}, and the spontaneous collapse proposal frequently
referred to as GRW \cite{Ghrr07,Frgg09}.  So far as the author is aware, none
of these interpretations has been developed to the point where it provides a
precise and coherent explanation of how a physical measurement process
functions, and in particular how the outcome (the traditional ``pointer
position'') of a measurement is related to the prior state of affairs which the
competent experimentalist designed his equipment to study. Thus one must await
further developments of these approaches by their proponents before assessing
whether they can provide insight into measurement contextuality.

The remainder of this paper is organized as follows.  Section~\ref{sct2} is a
review of quantum properties and variables as represented in a Hilbert
space. Measurements are the subject of Sec.~\ref{sct3}. They are discussed
using a quite general and fully quantum mechanical measurement model involving
no hidden variables, with the conclusion that an apparatus properly designed
to reveal the value that the observable $A$ had just before the measurement
took place will do just that, and this remains true if the apparatus also
measures a compatible (i.e., $[B,A]=0$) observable $B$ along with $A$.  It is
argued in Sec.~\ref{sct4} that this settles the contextuality question for
Hilbert space quantum mechanics.  The final Sec.~\ref{sct5} summarizes the
result and considers its relation to hidden variables approaches.  Some
comments on the single framework rule of histories quantum mechanics have been
placed in an appendix. 

\section{Quantum Properties and Variables}
\label{sct2}

We begin with a brief review of how the quantum properties of a system $s$
are represented using a Hilbert space $\HC_s$
which (as in Bell's original discussion) can be conveniently
assumed to be finite dimensional.  An \emph{observable} or \emph{quantum
variable} is represented by a Hermitian operator $A=A\ad$ on $\HC_s$ with
spectral representation
\begin{equation}
  A = \sum_\al a_\al P_\al,
\label{eqn1}
\end{equation}
where the $\{a_\al\}$  are the eigenvalues of $A$, and the $\{P_\al\}$ a
collection of projectors that form a \emph{decomposition of the identity} $I$:
\begin{equation}
  I= \sum_\al P_\al,\quad P_\al = P_\al\ad = P_\al^2,\quad
P_\al P_{\al'} = \dl_{\al\al'} P_\al.
\label{eqn2}
\end{equation}
We assume that $a_\al\neq a_{\al'}$ whenever $\al\neq\al'$, so the decomposition
$\{P_\al\}$ is uniquely determined by $A$.

A projector $P=P\ad= P^2$ projects onto a linear subspace of $\HC$, and there
is a one-to-one correspondence between projectors and subspaces, so one can
speak either of a collection of subspaces or of projectors; we shall generally
employ the latter.  Following von Neumann, Sec.~III.5 of \cite{vNmn32b}, we
think of a subspace, or equivalently the associated projector, as the
mathematical representation of a quantum \emph{property}, the quantum analog of
a collection of points $E$ in a classical phase space representing a classical
property such as ``the energy is less than 2 Joules.''  The quantum analog of a
single point in a classical phase space is a one-dimensional subspace (a ray)
consisting of complex multiples of any of its nonzero elements.

Thus for any observable $A$ there is a unique subspace $P_\al$ of $\HC$ where
$A$ takes on its eigenvalue $a_\al$; it consists of all the eigenvectors of
$A$ having this eigenvalue, together with the zero vector.  In classical
mechanics if a system is described by a point in the phase space inside the
set $E$ corresponding to some property we say that the system \emph{has} or
\emph{possesses} the property $E$.  Similarly, in quantum theory it makes
sense to say that if a system is described by a nonzero $\ket{\psi}$ in the
subspace $P_\al$, which is to say
\begin{equation}
  P_\al\ket{\psi} = \ket{\psi},
\label{eqn3}
\end{equation}
then the system has the property $P_\al$. (The reason for requiring that
$\ket{\psi}$ be nonzero is that the zero ket, which lies in every subspace, is
the quantum analog of the the empty set in a classical phase space: it
represents a proposition or property which is never true.)  If the quantum
system is described by a density operator $\rho$ such that
\begin{equation}
  P_\al\rho = \rho = \rho P_\al = P_\al\rho P_\al,
\label{eqn4}
\end{equation}
where any one of the equalities implies the others, it again makes sense to say
that the system has the property $P_\al$.  And when the system has the property
$P_\al$, the observable $A$ has the value $a_\al$. 

The decomposition $\{P_\al\}$ is the quantum analog of a coarse graining of
the classical phase space into a collection of nonoverlapping cells, and
represents a quantum \emph{sample space} in the sense employed in probability
theory: a set of mutually-exclusive properties (``events''), one and only one
of which is true, i.e., actually occurs, at a particular time.  Since $A$ is,
in effect, a real-valued function on this sample space it can be thought of as
a quantum analog of a ``random variable'' in probability theory, or of a
real-valued function on the classical phase space.  Thus it is appropriate to
refer to a Hermitian operator $A$ as a ``physical variable,'' in addition to
using the traditional term ``observable.''

The quantum-classical analogy cannot be pushed too far, since quantum theory
allows for noncommuting observables, whereas in classical mechanics all
physical variables (functions on the phase space) commute.  We shall call the
observables $A$ and $B$ \emph{compatible} when $[A,B]=0$; otherwise they are
(mutually) \emph{incompatible}.  Similarly, two projectors $P$ and $Q$ are
compatible if $[P,Q]=0$, in which case the product $PQ=QP$ is itself a
projector onto a subspace which represents the conjunction $P$ \AND\ $Q$ of
the properties $P$ and $Q$.  Otherwise they are incompatible, and defining the
conjunction $P$ \AND\ $Q$ is problematical.  In traditional quantum logic,
descending from the work of Birkhoff and von Neumann \cite{BrvN36}, $P$ \AND\
$Q$ is the property associated with the subspace which is the intersection of
the two subspaces corresponding to $P$ and $Q$.  This, however, requires the
use of new logical rules to avoid falling into difficulties---for a simple
example see Sec.~4.6 of \cite{Grff02c}.  But the main complaint about quantum
logic is that it has not resolved the major conceptual difficulties of quantum
mechanics \cite{Mdln05,Bccg09}.  In the histories approach one avoids these
difficulties by insisting that $P$ \AND\ $Q$ only makes sense, can only be
discussed, when the projectors commute. This is an example of the \emph{single
  framework rule}, a principle which can be used to resolve numerous quantum
paradoxes, as discussed in Chs.~20 to 25 of \cite{Grff02c}.  See the appendix
for some comments on this rule. 

Consider an observable $B$ that is compatible with $A$,
with a spectral representation
\begin{equation}
  B = \sum_\bt b_\bt Q_\bt,
\label{eqn5}
\end{equation}
where again the decomposition of the identity $\{Q_\bt\}$ is chosen so that
$b_\bt\neq b_{\bt'}$ whenever $\bt\neq \bt'$. The fact that $[A,B]=0$ implies
that $[P_\al,Q_\bt] = 0$ for every $\al$ and $\bt$, and consequently there is
a third decomposition of the identity $\{R_j\}$, the common refinement of
$\{P_\al\}$ and $\{Q_\bt\}$, consisting of all nonzero projectors of the form
$P_\al Q_\bt$, such that
\begin{equation}
  A = \sum_j a_j R_j,\quad B = \sum b_j R_j.
\label{eqn6}
\end{equation}
The $a_j$ in \eqref{eqn6} are of course eigenvalues of $A$, but now two or
more of them may have the same value, in contrast to \eqref{eqn1}, which is
the reason for using a distinct subscript; the same comment applies to
the $b_j$.  Because it consists of products of the form $P_\al Q_\bt$, the
decomposition $\{R_j\}$ is the smallest collection of projectors such that
both $A$ and $B$ can be written in the form \eqref{eqn6}.  If $\ket{\psi}$
belongs to a subspace $R_j$, then since $R_j$ is itself the product $P_\al
Q_\bt$ for some specific choice of $\al$ and $\bt$, it follows that if the
quantum system under discussion has the property $R_j$ it also has the
property $P_\al$ in the sense that \eqref{eqn3} is satisfied, and likewise the
(compatible) property $Q_\bt$.  This is obvious for the analogous situation of
a classical phase space, but in the quantum case it only makes sense when all
the projectors that we are considering, the $\{P_\al\}$, $\{Q_\bt\}$, and
$\{R_j\}$, commute with each other; as noted above there are logical
difficulties in the noncommuting case.

It should be noted that \emph{quantum physical variables (observables)} and
\emph{properties} as defined above apply not only to microscopic
quantities, but also to macroscopic objects, such as laboratory equipment.
There is no experimental evidence for a ``classical'' world to which quantum
principles do not apply.  Quantum theory is routinely used when discussing
properties of the solid materials out of which experimental apparatus is
constructed, the properties of stars, etc.  So far as we know at present, it
applies ``from the quarks to the quasars.''  To be sure, there are excellent
reasons why engineers discuss the motion of automobiles using classical
mechanics rather than quantum theory, but this is a practical matter, not one
of principle.  We now understand, at least in general terms, how in suitable
circumstances classical physics emerges as a good approximation to a more
fundamental quantum physics \cite{GMHr93,GMHr07}, and this justifies the fully
quantum mechanical description given below of a physical measurement carried
out by a macroscopic piece of apparatus.

\section{Measurements}
\label{sct3}

\subsection{Introduction}
\label{sbct3.1}

The most direct approach to determining whether quantum theory is or is not
contextual is to analyze the process that goes on in a quantum measurement.  A
measurement, as that term is ordinarily employed in physics, means determining
one or more properties possessed by a system at a time just preceding that at
which the measurement was carried out.  It is in this sense that experimental
particle physicists can speak, as they do, of detecting neutrinos arriving
from the sun, or measuring the energy of a particle shower that has come to
rest in their instrument.  The reason for referring to the time just
\emph{before} the measurement is carried out is that the measuring process
itself may change the property in question, and sometimes, as in the case of
photons, the object being measured actually disappears.  There is an important
difference between a measurement and a \emph{preparation} in which the
instrument and setup tells one something about the properties of the prepared
system \emph{after} the preparation takes place.  Considerable confusion is
created in discussions of quantum foundations, and in the minds of students
studying quantum theory for the first time, by the common failure to make a
clear distinction between measurement and preparation.  Part of this goes back
to von Neumann's treatment of measurements, Ch.~VI of \cite{vNmn32b}, in which
he employed a very special model that in effect combines a measurement with a
preparation.  Whatever the utility of this model when quantum theory was first
developed, greater clarity is possible nowadays if the distinction between
these processes is kept in mind \cite{Grff12}.

\subsection{Measurement model}
\label{sbct3.2}

We use a model of the measuring process in which the apparatus itself is a
fully quantum mechanical object subject to exactly the same physical laws as
the system being measured.  Let $\HC=\HC_s\ot\HC_m$ be the combined Hilbert
space of the system $s$ being measuring and the apparatus $m$. The latter can
include the environment of the apparatus and whatever else is relevant to the
description of the measuring process, enough so that we can consider $\HC$ the
Hilbert space of a closed system to which Schr\"odinger's equation
applies. That equation gives rise to a unitary time evolution operator%
\footnote{If the total system Hamiltonian $H$ is independent of time,
  $T(t_2,t_1) = e^{-i(t_2-t_1)H/\hbar}$.} %
$T(t_2,t_1)$ on $\HC$, where $t_1$ is a time just before the measurement takes
place and $t_2$ a time just after it is complete, with results indicated by
the position of the traditional pointer.  Let us assume that at the initial
time the system has the property $P_\al$ and the measuring apparatus (plus
environment, etc.) is in an initial macroscopic ``ready'' state corresponding
to the property or projector $M_0$ onto some subspace (typically of extremely
high dimension) of $\HC_m$. Thus at $t_1$ the combined system plus apparatus
has the property $P_\al\ot M_0$. (The reader who prefers a density operator
for the apparatus is welcome to employ that rather than $M_0$ in the following
discussion.)  Next we assume that there is a decomposition $\{\Pi_\al\}$ of
the identity $I$ for $\HC$ in which the index $\al$ labels macrostates
corresponding to different possible positions of the pointer indicating
the measurement outcome, and in addition to the values in \eqref{eqn1} there
is another, say $\al=0$, such that
\begin{equation}
  \Pi_0 = I - \sum_{\al\neq 0} \Pi_\al.
\label{eqn7}
\end{equation}
The projector $\Pi_0$ allows for all other possibilities; e.g., the apparatus
was not properly set up in the first place.  Next we assume that the
unitary dynamics is such that if we define
\begin{equation}
  V_\al :=T(t_2,t_1) \blp P_\al\ot M_0\brp T(t_1,t_2),
\label{eqn8}
\end{equation}
it is the case that
\begin{equation}
  \Pi_{\al'} V_{\al} = \dl_{\al\al'} V_\al.
\label{eqn9}
\end{equation}
In words, if the system has property $P_\al$ at time $t_1$, then the
time-evolved property $V_\al$ is contained in the subspace onto which
$\Pi_\al$ projects, meaning that the pointer is in the position $\al$.  The
$\dl_{\al\al'}$ implies that the system property $P_\al$ at $t_1$ and the
pointer at $\Pi_\al$ at $t_2$ are perfectly correlated: from the latter one
can infer the former, and vice versa.  Notice that the $\Pi_\al$ are
projectors on $\HC$, not on $\HC_m$ alone, and are associated with pointer
positions. They say nothing at all about the final state of the measured
system $s$, which may not even exist; e.g., as when a photon is absorbed by the
detector. We are, as noted earlier, discussing a
measurement, not a preparation.

The reader unfamiliar with this type of measurement model may find it helpful
to consider the special case in which $M_0=\dya{M_0}=[M_0]$, where we use the
abbreviation $[\psi]=\dya{\psi}$ for the projector onto the 
ray containing $\ket{\psi}$, meaning the  apparatus is in an
initial pure state $\ket{M_0}$, and the system to be measured is in a pure
state $\ket{a_\al}$ for some $\al$, so $P_\al=[a_\al]$. Then if we define
\begin{equation}
  \ket{V_\al} := T(t_2,t_1) \blp \ket{a_\al} \ot \ket{\Psi_1}\brp,
\label{eqn10}
\end{equation}
$V_\al$ in \eqref{eqn8} is $[V_\al]$, and \eqref{eqn9} is equivalent to
\begin{equation}
  \Pi_{\al'} \ket{V_\al} = \dl_{\al\al'} \ket{V_\al},
\label{eqn11}
\end{equation}
which is analogous to \eqref{eqn3}: when $\al=\al'$ the total system has the
property that the pointer is in position $\al$. 

Obviously this is an idealized description of an actual laboratory
measurement, but it is nonetheless accurate in representing the essential
features of the latter, in somewhat the same way in which a diagram of a
telescope showing rays entering the aperture before bouncing off mirrors and
exiting through a lens represents the essential idea of how it operates.  It
is the task of the competent experimentalist to arrange a piece of apparatus
so that the transition from $t_1$ to $t_2$ takes place in way such that
\eqref{eqn9} is satisfied.  He can test whether or not it functions according
to design by making repeated runs in which the apparatus $m$ always starts off
in the ready state $M_0$, and the system $s$ is prepared in one of the states
$P_\al$, where $\al$ varies from run to run. If the final outcome (pointer
position) always corresponds to the known initial state of the system, this
indicates that \eqref{eqn9} is satisfied.

\subsection{Superposition states}
\label{sbct3.3}

What will occur if the experimenter prepares an initial state
\begin{equation}
  \ket{\psi} = (\ket{a_1} + \ket{a_2})/\st,
\label{eqn12}
\end{equation}
which is a superposition corresponding to distinct eigenvalues of $a_1$ and
$a_2$ of $A$ and then carries out a measurement? The answer is that the
outcome will differ from run to run.  Sometimes the pointer (or its electronic
counterpart in a modern laboratory) will have the property corresponding to
$\Pi_1$, and sometimes that corresponding to $\Pi_2$, with roughly half the
runs indicating one and half indicating the other.  We agree with Appleby
\cite{Appl05} that probabilities can only come from probabilities, so in order
to obtain the probabilities of standard quantum mechanics one must introduce a
probabilistic axiom.  The simplest is the Born rule, which for present
purposes can be worded as follows.  Suppose that at $t_1$ we assign a property
$\hat P$ to $s$, where $\hat P$ projects on any subspace of $\HC_s$, and
define
\begin{equation}
  \hat V :=T(t_2,t_1) \blp \hat P\ot M_0\brp T(t_1,t_2).
\label{eqn13}
\end{equation}
At $t_2$ we are interested in
a collection of properties of the apparatus represented by the $\Pi_\al$,
which form a sample space of mutually-exclusive properties, to which
Born's rule assigns conditional probabilities
\begin{equation}
  \Pr(\Pi_\al\vb \hat P) = \Tr(\Pi_\al \hat V)/\Tr(\hat V),
\label{eqn14}
\end{equation}
where the denominator on the right side normalizes the probability.  If
$\hat P=P_{\al'}$, then $\Pr(\Pi_\al\vb \hat P) = \dl_{\al\al'}$, consistent
with the result in Sec.~\ref{sbct3.2}, but now expressed in the form of a
probability.  

If, on the other hand, one sets $\hat P = [\psi]$, the projector onto
$\ket{\psi}$ in \eqref{eqn12}, the $\hat V$ resulting from the time evolution
in \eqref{eqn13} will be a projector onto a coherent superposition of
macroscopically-distinct pointer positions.  As long as we are only concerned
with calculating the probabilities of the $\Pi_\al$, $\hat V$ in \eqref{eqn14}
serves simply as a calculational tool, a pre-probability in the terminology of
Sec.~9.4 of \cite{Grff02c}, and does not count as a physical property.  If, on
the other hand, one wishes to consider $\hat V$ a physical property, a member
of the decomposition of the identity formed by the projectors $\hat V$ and
$I-\hat V$, the Born rule assigns a probability $\Pr(\hat V\vb \hat P)=1$: the
property $\hat V$, a form of ``Schr\"odinger cat'', is certain to occur at
$t_2$.  How is this to be reconciled with the previous assignment of
probabilities to $\Pi_1$ and $\Pi_2$?  Observe that just as $\hat P=[\psi]$
does not commute with either $P_1$ or $P_2$, $\hat V$ does not commute with
$\Pi_1$ and $\Pi_2$, so it makes no sense---it violates the single framework
rule---to put all three together into a single discussion.  There is no law of
nature which identifies the correct framework to be used for a quantum
description.  However, if one wants to think of the process under discussion
\emph{as a measurement} with some outcome, the way a physicist would wish to
talk about it, the useful framework for this purpose is the collection of
properties $\{\Pi_\al\}$, not the framework $\{\hat V, I-\hat V\}$ which makes
it impossible to discuss outcomes. 

Thus if one employs the unitarily evolved state $\hat V$
(Schr\"odinger's cat) as a physical property at time $t_2$ there is no way in
which one can assert that the measurement has had a definite outcome, that the
apparatus pointer is in a particular state. 
In this respect the histories approach is distinctly different from the Everett
or many-worlds interpretation with its insistence that ``the wave function'',
in this instance the state onto which $\hat V$ projects, represents fundamental
physical reality.  From the histories point of view the difficulties which
many-worlds advocates have in explaining how ordinary macroscopic physics can
be consistent with their perspective is not unrelated to the fact that they are
seeking to assign simultaneous reality to properties which in the quantum
Hilbert space are represented by incompatible projectors.  (For further remarks
on the single framework rule see the appendix, and for more details the
discussion in Sec.~2.3 of \cite{Grff12}.)

But in addition to getting rid of the ghost of Schr\"odinger's cat, we still
need to show that the measurement apparatus actually carries out a
\emph{measurement}; i.e., the outcome pointer position is properly correlated
with a previous property of the measured system. For this purpose we need an
extension of Born's rule that allows probabilities to be assigned to a closed
quantum system at three or more times, and this in turn requires the use of
consistent (or decoherent) families of histories. As this topic is discussed
in considerable detail and with numerous examples in \cite{Grff02c}, with more
abbreviated treatments in \cite{Grff11b,Hhnb10}, we omit details and give
only the essential results as they apply in the present situation.

To this end it is convenient to introduce a time $t_0$ slightly earlier than
$t_1$, but still later than the completion of the preparation procedure that
results in the state $\ket{\psi}$ in \eqref{eqn12}. 
  We assume that no significant time
development goes on during the short interval from $t_0$ to $t_1$, and 
consider a family of histories
\begin{equation}
  [\psi]\ot M_0 \od \{P_\al\} \od \{\Pi_\al\},
\label{eqn15}
\end{equation}
where the notation is to be interpreted as follows.  The symbol $\od$
separates situations at the different times; it is the counterpart of ``$<$''
in $t_0 < t_1 < t_2$.  The initial state at time $t_0$ is a projector
$[\psi]\ot M_0$ with $[\psi] = \dya{\psi}$ corresponding to the superposition
$\ket{\psi}$ in \eqref{eqn12}.  At $t_1$ we consider the collection of
properties $\{P_\al\}$ making up the decomposition in \eqref{eqn2}.
(Following the usual convention of physicists, the symbol $P_\al$ should be
understood as $P_\al\ot I_m$ when used in a context in which the full Hilbert
space is $\HC_s\ot\HC_m$.) Similarly at time $t_2$ the properties under
consideration are the pointer positions $\{\Pi_\al\}$.  An extension of the
Born rule allows the assignment of probabilities to the different histories,
each history involving one of the properties in \eqref{eqn15} at each time,
provided appropriate consistency (or decoherence) conditions are satisfied.
In the present case consistency is a consequence of \eqref{eqn9}, and the
extended Born rule assigns zero probability to each history except for the
pair
\begin{equation}
  [\psi]\ot M_0\od P_1\od \Pi_1,\quad
  [\psi]\ot M_0\od P_2\od \Pi_2.
\label{eqn16}
\end{equation}
To each of these it assigns a probability of 1/2.  As a consequence the
conditional probability of the property $P_1$ at time $t_1$ given the pointer
position $\Pi_1$ at $t_1$ is equal to 1, and vice versa; and the same for
$P_2$ and $\Pi_2$.  Of course there is nothing special about the superposition
used in \eqref{eqn12}; any other would have led to a similar result, which is
just the conclusion we reached earlier in Sec.~\ref{sbct3.2}: there is an
exact, one-to-one correspondence between later pointer positions and earlier
properties of the system to be measured.

The family of histories in \eqref{eqn15} is not the only possible choice for a
consistent quantum description.  One could instead employ
\begin{equation}
  [\psi]\ot M_0 \od \{[\psi], I-[\psi]\} \od \{\Pi_\al\},
\label{eqn17}
\end{equation}
with a different choice of properties for $s$ at the intermediate time $t_1$.
Indeed, \eqref{eqn17} corresponds fairly closely to a systematic quantum
description of the narrative students are taught in textbooks, in which there
is nothing but unitary time development up to the instant at which a
measurement takes place, when by fiat a ``collapse'' takes place in which
pointer positions somehow emerge out of a superposition fog.  There is nothing
wrong with \eqref{eqn17} in terms of fundamental quantum mechanics, but it is
not suitable for discussing a \emph{measurement}, in contrast to a
preparation, since the properties which the measurement was designed to reveal
cannot be discussed using this family: the $P_\al$ will (in general) not
commute with $[\psi]$.  Thus the choice between \eqref{eqn15} and
\eqref{eqn17} is not a matter of fundamental quantum theory, where both are
equally satisfactory, but instead a matter of utility: only the former allows
one to discuss a measurement as a measurement.  (For additional remarks on the
relationship of incompatible families, see \cite{Grff12}.)

In summary, a properly constructed quantum measurement apparatus does just
what it was designed to do, measure properties of a system corresponding to a
particular decomposition of the identity. It will do this for a system
initially prepared in one of the states it was designed to measure, and for
one initially prepared in a superposition of those states.  To discuss the
latter one needs to employ probabilities, in particular the (extended) Born
rule, both in order to have a measurement pointer with a well-defined
position, and to show that this position is appropriately correlated with one
of the properties the apparatus was designed to measure.  For details omitted
from the preceding discussion the reader is referred to the extensive
treatment of measurements in \cite{Grff02c}, as well as the shorter
discussions in \cite{Hhnb10,Grff11b,Grff12}.

\subsection{Measuring compatible observables}
\label{sbct3.4}

Next consider a situation in which two observables, $A$ associated with the
decomposition $\{P_\al\}$ and $B$ with $\{Q_\bt\}$ commute with each other. As
discussed in Sec.~\ref{sct2} there is a decomposition $\{R_j\}$ of the
identity, consisting of products $P_\al Q_\bt$, which allows both $A$ and $B$
to be written in the form \eqref{eqn6}. To determine the values of both of
them it suffices (and is also necessary) to carry out a measurement which
determines which property $R_j$ is possessed by the system.  To do such a
measurement we follow the same strategy as in Sec.~\ref{sbct3.2}, using an
apparatus $m'$ in an initial state $M'_0$, and a decomposition $\{\Xi_j\}$
corresponding to pointer positions replacing the $\{\Pi_\al\}$. Use operators
\begin{equation}
  W_j:= T(t_2,t_1)\left( R_j\ot M'_0\right) T(t_1,t_2)
\label{eqn18}
\end{equation}
in place of the $V_\al$, and replace \eqref{eqn9} with
\begin{equation}
  \Xi_j W_{j'} = \dl_{jj'} W_j.
\label{eqn19}
\end{equation}
The argument that each outcome $\Xi_j$ reflects a prior property $R_j$ 
is then the same as that used earlier.

A particular $P_\al$ may be equal to one of the $R_j$, or it may be the sum of
two or more. Suppose, for example, that $P_1=R_3+R_4$. If the
experimenter prepares the system with a property $P_1$ then it may have
neither of the properties $R_3$ or $R_4$. But it will still be the case that
as a consequence of the linearity of \eqref{eqn19},
\begin{equation}
  (\Xi_3+\Xi_4) V_1 = V_1,
\label{eqn20}
\end{equation}
where $V_1$ as defined in \eqref{eqn8}.  Hence, again invoking the Born rule,
the measurement outcome will be one of the pointer positions corresponding to
$\Xi_1$ or $\Xi_2$, varying randomly from run to run.  These two outcomes are
uniquely associated with the property $P_1$ in that they cannot occur if the
initial property is $P_\al$ with $\al\neq 1$. Thus either outcome implies that
the system earlier possessed the property $P_1$.  Therefore this new procedure
remains valid for measuring $A$, while at the same time, by the same sort of
argument, it provides a measurement of $B$.

\section{Noncontextuality}
\label{sct4}

The preceding argument showing that a joint measurement of an observable $A$
and a compatible observable $B$ using a properly constructed apparatus
apparatus $m'$ will reveal a property $P_\al$, corresponding to $A=a_\al$, at a
time before the measurement took place, is independent of the choice of $B$ as
long as $[A,B]=0$.  Consequently, if the apparatus, call it $m''$, is instead
designed to measure $A$ and a different observable $C$, where $[A,C]=0$ but
$[B,C]\neq 0$, the connection between the measurement outcome and the prior
property $P_\al$ corresponding to $a_\al$ is exactly the same.  And
since---always assuming a properly constructed apparatus---the outcome so far
as the value of $A$ is concerned is in both cases determined by the property
$P_\al$ the system $s$ possessed \emph{before} the measurement took place, it
can make no difference which apparatus, $m'$ or $m''$, is used to carry out the
measurement.  The result will be exactly the same given that the original
quantum property of $s$ was the same.  The choice between apparatuses $m'$ and
$m''$ makes no difference for properties that correspond to some eigenvalue
$a_\al$ of $A$.

Another way of thinking about the matter is to suppose that there is a single
measurement apparatus, but it is supplied with a switch that allows it to be
changed from a kind that can measure both $A$ and $B$ to a kind that can
measure both $A$ and $C$, and the setting of this switch is determined by a
coin toss that takes place just prior to the arrival of $s$ at the measuring
device.  (This can be modeled, if one chooses, using a closed quantum system by
employing a quantum coin toss, of the sort discussed in Sec.~19.2 of
\cite{Grff02c}.)  Let us suppose that in some particular case the switch
setting results in a measurement of $A$ and $B$, with an outcome from which one
can, with probability 1, infer that $s$ had, say the property $P_3$ (i.e.,
$\al=3$) at a time $t_1$ before the quantum coin was tossed.  We can then ask
the counterfactual question: what \emph{would} have been the result of the
measurement \emph{if} the coin toss had resulted in the measurement of $A$ and
$C$, rather than $A$ and $B$?

Analyzing counterfactual questions is a bit tricky even in situations not
entangled with quantum mysteries.  There is, however, an approach, see
Sec.~19.4 of \cite{Grff02c} and later chapters for various applications, which
seems to work well in the quantum case, and has withstood (in the author's
opinion; the reader is invited to examine the references) a recent
attack by Stapp \cite{Stpp12}, with response in \cite{Grff12b}.  It amounts to
starting with an event at a later time in the actual world $W$, tracing it
backwards in time to some property, referred to as a ``pivot,'' that $W$ has
in common with the counterfactual world $W'$, and then using quantum
theory to reason forwards in time to a later property of $W'$.  In the present
situation---we leave the details of the analysis to the interested reader---if
the property $P_3$ at time $t_1$ is used as a pivot, the conclusion is that if
$A$ and $C$ had been measured, the result so far as the property $A$ is
concerned would have been exactly the same, that is, a pointer position from
which the earlier $P_3$ could have been inferred with probability 1.

\section{Discussion}
\label{sct5}

The preceding argument establishes the noncontextuality of quantum mechanics
when the measured quantities are quantum properties associated with subspaces
of the Hilbert space. The essential idea is that apparatus competently
designed and constructed to reveal a property which the system of interest had
just before the measurement takes place will do exactly that, and it is
irrelevant what other properties, if any, the apparatus has been designed to
measure.  (Of course, incompatible properties cannot be simultaneously
measured, because in Hilbert space quantum mechanics there is nothing there to
be measured.) 

Given this fairly straightforward way of demonstrating that quantum mechanics
is noncontextual, it is natural to ask why the topic of contextuality still
seems to attract considerable attention, even leading to proposals for
experimental tests.  In this connection it is helpful to examine the recent
paper by Hermens \cite{Hrmn11} already referred to in Sec.~\ref{sct1}.  He
gives a list of four statements which because of the Kochen-Specker theorem
cannot all be true:
\begin{quote}
  \textit{QM} (Quantum mechanics): Every observable $\AC$ can be associated
  with a self-adjoint operator $A$ on some Hilbert space $\HC$. The result of
  a measurement of $\AC$ is an element of the spectrum $\sg(A)$ of
  $A$. Observables whose corresponding operators commute can be measured
  simultaneously. Moreover, if there is a functional relationship between the
  associated operators, this relation is preserved in the measurement results.

  \textit{Re} (Realism): Every observable $\AC$ possesses a certain (real)
  value $\lm(A)$ at all times.

  \textit{FM} (Faithful measurement): A measurement of an observable $\AC$ at
  a certain time reveals the value $\lm(\AC)$ possessed by that observable at
  that time.

  \textit{CP} (Correspondence principle): There is a bijective correspondence
  between observables and self-adjoint operators.
\end{quote}

To connect these statements with our previous discussion, note that we have
used the same symbol $A$ for a physical variable or observable, and the
corresponding quantum operator, which is to say we assumed the validity of
Hermens' \textit{CP} from the very outset: ``observable'' is just the name
associated with self-adjoint or Hermitian (equivalent for finite-dimensional
$\HC$) operators in the usual textbook approach to quantum theory.  Remember
that we are employing Hilbert space quantum mechanics, not a Hilbert space
augmented with hidden variables, and from this perspective a one-to-one
association of self-adjoint operators with quantum observables, or,
equivalently, projectors with quantum properties, is entirely reasonable.  We
also agree with Hermen's \textit{QM}.%
\footnote{A functional relationship between operators implies a functional
  relationship between the sets of eigenvalues when both are expressed using a
  common decomposition of the identity. E.g., one could imagine that there is a
  (real-valued) function $f(x)$ such that for every $j$ in \eqref{eqn6} $b_j =
  f(a_j)$. What the measurement actually reveals is the prior property $R_j$,
  but if one uses it to infer the corresponding eigenvalues the functional
  relationship between them will, of course, be satisfied.} %
Furthermore, our measurement model, with $\lm(\AC)$ an eigenvalue of $A$,
satisfies \textit{FM}, though with the qualification that the measured property
was possessed by the system at a time just before the measurement took place.
There is, to be sure, a subtle point involved in the notion that an observable
can ``possess'' a value, especially since the observable $A$ can in general
occur in (i.e., the corresponding decomposition of the identity as in
\eqref{eqn1}) a variety of different frameworks, including some that are
incompatible with others.  See the additional remarks in the appendix

The way our approach avoids any conflict with the
Kochen-Specker theorem is by denying \textit{Re}. The claim that \emph{every}
observable possesses a value at every time is, indeed, inconsistent with a
representation of quantum properties by subspaces of a Hilbert space.
Consider, for example, a spin-half particle. There are distinct rays in the
two-dimensional Hilbert space corresponding to $S_x=+1/2$ (in units of
$\hbar$) and to $S_x=-1/2$; also rays corresponding to $S_z=+1/2$ and
$S_z=-1/2$, but there is no ray that can represent a simultaneous value of
$S_x$ \emph{and}  $S_z$.  Students are told, correctly, that $S_x$ and $S_z$
cannot be measured simultaneously, and they ought to be told that the reason
for this is that there is nothing there to be measured. The projectors
corresponding to $S_z$ do not commute with the projectors corresponding to
$S_x$, and once one has accepted the connection between quantum properties and
Hilbert subspaces proposed by von Neumann it makes no sense to speak of a spin
half system in which, for example, $S_x=+1/2$ at the same time that
$S_z=+1/2$.  For further discussion of these matters see Sec.~4.6 of
\cite{Grff02c}.  For additional comments on how the histories approach is
related to the Kochen-Specker result, see the appendix.

Consequently, three of Hermens' four statements (perhaps slightly modified)
are in agreement with Hilbert space quantum mechanics, but \textit{Re} is
not. It is somewhat odd that this particular principle should be identified
with \emph{realism}, since at the present time all available experimental
evidence is in accord with Hilbert space quantum mechanics, and not with
classical physics when the two disagree.  If a Hilbert space provides the
appropriate mathematics to describe everything from the quarks to the quasars,
where in the real world, the one we live in, is there any part that satisfies
the condition of ``realism'' given by \textit{Re}?  It would be much less
confusing if whenever ``realism'' were used in this way the adjective
``classical'' were prepended.  The hidden variables of typical hidden variable
theories are \emph{classical} hidden variables, and it is for this reason that
the attempt to use them for interpreting quantum theory has given rise to
numerous conflicts with the latter.

Not only have classical hidden variables given rise to misleading notions
about the contextuality of quantum measurements, they have also contributed to
the misleading idea that quantum mechanics is nonlocal in the sense that
actions at one point in spacetime have a mysterious nonlocal influence on what
goes on at other, spatially separated, points in spacetime, contrary to
special relativity.  For a demonstration that Hilbert space quantum theory
contains no such influences, see \cite{Grff11, Grff11b}.  It is thus
surprising that classical hidden variables continue to receive so much
attention in discussions of the foundations of quantum mechanics when they are
of such little use.  Perhaps it is time to pay serious attention to the
mathematical properties of Hilbert space, and the representation of physical
properties by means of its subspaces.

\section*{Acknowledgments}

The author thanks R. Hermens for helpful correspondence, and two anonymous
referees for their careful reading of the text and suggestions for clarifying
it.  The research described here received support from the National Science
Foundation through Grant PHY-1068331.  

\appendix

\section{Appendix. The Single Framework Rule}
\label{apx1}

The histories approach to Hilbert space quantum mechanics adopts a precise
formulation of quantum probabilities constructed in the following way.  A
probabilistic theory requires a \emph{sample space} of mutually exclusive
possibilities or events, one and only one of which is true (or occurs), an
\emph{event algebra}, and some way of assigning probabilities to the latter.  A
quantum sample space in the histories approach is always a projective
decomposition of the identity, as in \eqref{eqn2}. In the case of
time-dependent stochastic quantum processes the same idea is used, but the
decomposition now refers to projectors on the \emph{history Hilbert space},
ordered sequences of events at discrete times, which constitute the elementary
histories; for details see Ch.~8 of \cite{Grff02c}. In either case the event
algebra is constituted by sums of collections of projectors from this sample
space, and probabilities are assigned to compound events in the usual fashion,
as sums of the probabilities assigned to the individual events that they
contain. Either the original sample space or the event algebra is referred to
as a quantum \emph{framework}.

The \emph{single framework rule} asserts that any sort of discussion of the
quantum system be carried out in some framework of the sort just discussed,
which is typically chosen because it has some events which are interesting for
some reason or another.  The physicist is free to choose any framework he
pleases for describing the world; what the single framework rule prohibits is
\emph{combining} frameworks. If two (or more) frameworks are \emph{compatible}
in the sense that all the projectors in one commute with all the projectors in
the other one can get around this prohibition by using the common refinement, a
sample space formed by nonzero products of the projectors in the two original
sample spaces. However, if the frameworks are \emph{incompatible}, which is to
say some projectors from one do not commute with some projectors from the
other, a common refinement is no longer possible, and single framework rule
forbids using logical reasoning which bridges the two.

The well-known Kochen-Specker paradox is constructed precisely by forming a
bridge, or one might better say bridges, between incompatible frameworks in
such a way that one eventually ends up with a contradiction.  The histories
approach disposes of the paradox by declaring the bridging invalid.  For more
details, including a refutation of the claim by Bassi and Ghirardi
\cite{BsGh99,BsGh00} that Kochen-Specker invalidates the histories approach, see
\cite{Grff00,Grff00b} and the discussion in Ch.~22 of \cite{Grff02c}.

An additional aspect of the single framework rule (or one might say an
extension) appears when considering quantum time dependence.  In order to
extend the Born rule for probabilities in a consistent way to histories
involving three or more times it is necessary to impose \emph{consistency
  conditions}, see Ch.~10 of \cite{Grff02c}.  There are examples, of which the
best known is the three-box paradox of Aharonov and Vaidman \cite{AhVd91}---see
the discussion in Sec.~22.5 of \cite{Grff02c}---where combining distinct sample
spaces of histories is not prohibited by noncommuting projectors, but in which
consistency conditions are not longer satisfied by the combination. The
(extended) single framework rule then rules out contradictions of the sort
claimed by Kent \cite{Knt97} and refuted in \cite{GrHr98}.

It is important to note that the single-framework rule does \emph{not} state
that there is only one framework which can be used for a valid quantum
description of a situation.  The quantum physicist is free to choose \emph{any}
framework, consistent with the Hilbert space structure of quantum mechanics
(and, in the case of histories, satisfying consistency conditions if the
extension of the Born rule is to be used to assign probabilities), in order to
describe a quantum system.  In some frameworks a particular observable $A$
may possess some value, while in other frameworks it may not.  The existence of
the latter does not preclude the possibility or the validity of a framework
which include the former. The single framework rule is not a restriction on the
\emph{use} of frameworks; it is instead a prohibition against \emph{combining}
incompatible frameworks. 

Given the possibility, and indeed validity, of employing a variety of different
frameworks the reader might suppose that the histories approach will inevitably
produce contradictions.  That it does not do so can be seen from the following
consistency result discussed in Ch.~16 of \cite{Grff02c}.  Suppose one wishes
to infer from some initial---in the sense that they form the beginning of an
argument, not that they are necessarily earlier in time---data that at
such-and-such a time a quantum system possesses some property, say a physical
variable or observable $A$ has some value.  This is done by constructing a
framework which includes both the data and the observable $A$ (i.e., the
corresponding decomposition of the identity, as in \eqref{eqn1}), and then
applying the laws of quantum mechanics to reach a conclusion, which in general
will be probabilistic.  Typically there are a variety of frameworks available
for this task, some of which are incompatible with others.  The consistency
result is that one will reach the same (probabilistic) conclusion in any one of
these frameworks. There will also be frameworks in which either the data or the
observable $A$ are not represented; these obviously cannot be used to draw such
a conclusion, but their existence as (possible) quantum descriptions does not
invalidate a conclusion drawn from a framework that includes both the data and
$A$.

\end{document}